\input harvmac

\noblackbox

%Macro for table
\input tables
\newcount\tabno
\tabno=0
\def\tablabel#1{\global\advance\tabno by 1\xdef#1{\the\tabno}}
\input epsf

% Blackboard Bold
 \font\blackboard=msbm10 % scaled \magstep1
 \font\blackboards=msbm7 \font\blackboardss=msbm5
 \newfam\black \textfont\black=\blackboard
 \scriptfont\black=\blackboards \scriptscriptfont\black=\blackboardss
 \def\Bbb#1{{\fam\black\relax#1}}

% Macro YT
\def\cN{{\cal N}}
\def\cO{{\cal O}}
\def\bZ{{\Bbb Z}}
\def\bC{{\Bbb C}}
\def\bR{{\Bbb R}}
\def\vol{{\rm vol}\,}
\def\tr{\mathop{\rm tr}}

\lref\CheungID{
  Y.~K.~Cheung, O.~J.~Ganor and M.~Krogh,
  ``Correlators of the global symmetry currents of 4D and 6D  superconformal
  theories,''
  Nucl.\ Phys.\  B {\bf 523}, 171 (1998)
  [arXiv:hep-th/9710053].
  %%CITATION = NUPHA,B523,171;%%
}

%\DouglasJS
\lref\DouglasJS{
  M.~R.~Douglas, D.~A.~Lowe and J.~H.~Schwarz,
  ``Probing F-theory with multiple branes,''
  Phys.\ Lett.\  B {\bf 394}, 297 (1997)
  [arXiv:hep-th/9612062].
  %%CITATION = PHLTA,B394,297;%%
}

%\AharonyEN
\lref\AharonyEN{
  O.~Aharony, J.~Sonnenschein, S.~Yankielowicz and S.~Theisen,
  ``Field theory questions for string theory answers,''
  Nucl.\ Phys.\  B {\bf 493}, 177 (1997)
  [arXiv:hep-th/9611222].
  %%CITATION = NUPHA,B493,177;%%
}

%\BanksNJ
\lref\BanksNJ{
  T.~Banks, M.~R.~Douglas and N.~Seiberg,
  ``Probing F-theory with branes,''
  Phys.\ Lett.\  B {\bf 387}, 278 (1996)
  [arXiv:hep-th/9605199].
  %%CITATION = PHLTA,B387,278;%%
}

%\SenVD
\lref\SenVD{
  A.~Sen,
  ``F-theory and Orientifolds,''
  Nucl.\ Phys.\  B {\bf 475}, 562 (1996)
  [arXiv:hep-th/9605150].
  %%CITATION = NUPHA,B475,562;%%
}

%\MorrisonPP
\lref\MorrisonPP{
  D.~R.~Morrison and C.~Vafa,
  ``Compactifications of F-Theory on Calabi--Yau Threefolds -- II,''
  Nucl.\ Phys.\  B {\bf 476}, 437 (1996)
  [arXiv:hep-th/9603161].
  %%CITATION = NUPHA,B476,437;%%
}

%\MorrisonNA
\lref\MorrisonNA{
  D.~R.~Morrison and C.~Vafa,
  ``Compactifications of F-Theory on Calabi--Yau Threefolds -- I,''
  Nucl.\ Phys.\  B {\bf 473}, 74 (1996)
  [arXiv:hep-th/9602114].
  %%CITATION = NUPHA,B473,74;%%
}

%\AnselmiAM
\lref\AnselmiAM{
  D.~Anselmi, D.~Z.~Freedman, M.~T.~Grisaru and A.~A.~Johansen,
  ``Nonperturbative formulas for central functions of supersymmetric gauge
  theories,''
  Nucl.\ Phys.\  B {\bf 526}, 543 (1998)
  [arXiv:hep-th/9708042].
  %%CITATION = NUPHA,B526,543;%%
}

%\VafaXN
\lref\VafaXN{
  C.~Vafa,
  ``Evidence for F-Theory,''
  Nucl.\ Phys.\  B {\bf 469}, 403 (1996)
  [arXiv:hep-th/9602022].
  %%CITATION = NUPHA,B469,403;%%
}

%\SeibergAJ
\lref\SeibergAJ{
  N.~Seiberg and E.~Witten,
  ``Monopoles, duality and chiral symmetry breaking in $\cN=2$ supersymmetric
  QCD,''
  Nucl.\ Phys.\  B {\bf 431}, 484 (1994)
  [arXiv:hep-th/9408099].
  %%CITATION = NUPHA,B431,484;%%
}

\lref\ArgyresCN{
  P.~C.~Argyres and N.~Seiberg,
  ``S-duality in $\cN=2$ supersymmetric gauge theories,''
  arXiv:0711.0054 [hep-th].
  %%CITATION = ARXIV:0711.0054;%%
}

\lref\FayyazuddinFB{
  A.~Fayyazuddin and M.~Spali\'nski,
  ``Large $N$ superconformal gauge theories and supergravity orientifolds,''
  Nucl.\ Phys.\  B {\bf 535}, 219 (1998)
  [arXiv:hep-th/9805096].
  %%CITATION = NUPHA,B535,219;%%
}

\lref\AharonyXZ{
  O.~Aharony, A.~Fayyazuddin and J.~M.~Maldacena,
  ``The large $N$ limit of $\cN = 2,1$ field theories from three-branes in
  F-theory,''
  JHEP {\bf 9807}, 013 (1998)
  [arXiv:hep-th/9806159].
  %%CITATION = JHEPA,9807,013;%%
}

\lref\AharonyRZ{
  O.~Aharony, J.~Pawe\l czyk, S.~Theisen and S.~Yankielowicz,
  ``A note on anomalies in the AdS/CFT correspondence,''
  Phys.\ Rev.\  D {\bf 60}, 066001 (1999)
  [arXiv:hep-th/9901134].
  %%CITATION = PHRVA,D60,066001;%%
}

\lref\ArgyresXN{
  P.~C.~Argyres, M.~Ronen Plesser, N.~Seiberg and E.~Witten,
  ``New $\cN=2$ Superconformal Field Theories in Four Dimensions,''
  Nucl.\ Phys.\  B {\bf 461}, 71 (1996)
  [arXiv:hep-th/9511154].
  %%CITATION = NUPHA,B461,71;%%
}
\lref\EguchiVU{
  T.~Eguchi, K.~Hori, K.~Ito and S.~K.~Yang,
  ``Study of $\cN=2$ Superconformal Field Theories in $4$ Dimensions,''
  Nucl.\ Phys.\  B {\bf 471}, 430 (1996)
  [arXiv:hep-th/9603002].
  %%CITATION = NUPHA,B471,430;%%
}
%\FreedTG
\lref\FreedTG{
  D.~Freed, J.~A.~Harvey, R.~Minasian and G.~W.~Moore,
  ``Gravitational anomaly cancellation for M-theory fivebranes,''
  Adv.\ Theor.\ Math.\ Phys.\  {\bf 2}, 601 (1998)
  [arXiv:hep-th/9803205].
  %%CITATION = 00203,2,601;%%
}
%\HarveyBX
\lref\HarveyBX{
  J.~A.~Harvey, R.~Minasian and G.~W.~Moore,
  ``Non-abelian tensor-multiplet anomalies,''
  JHEP {\bf 9809}, 004 (1998)
  [arXiv:hep-th/9808060].
  %%CITATION = JHEPA,9809,004;%%
}
\lref\MinahanFG{
  J.~A.~Minahan and D.~Nemeschansky,
  ``An $\cN = 2$ superconformal fixed point with $E_6$ global symmetry,''
  Nucl.\ Phys.\  B {\bf 482}, 142 (1996)
  [arXiv:hep-th/9608047].
  %%CITATION = NUPHA,B482,142;%%
}
\lref\MinahanCJ{
  J.~A.~Minahan and D.~Nemeschansky,
  ``Superconformal fixed points with $E_n$ global symmetry,''
  Nucl.\ Phys.\  B {\bf 489}, 24 (1997)
  [arXiv:hep-th/9610076].
  %%CITATION = NUPHA,B489,24;%%
}

%\KimEZ
\lref\KimEZ{
  H.~J.~Kim, L.~J.~Romans and P.~van Nieuwenhuizen,
  ``The Mass Spectrum Of Chiral $\cN=2$ $D=10$ Supergravity On $S^5$,''
  Phys.\ Rev.\  D {\bf 32}, 389 (1985).
  %%CITATION = PHRVA,D32,389;%%
}

\lref\ArgyresJJ{
  P.~C.~Argyres and M.~R.~Douglas,
  ``New phenomena in $SU(3)$ supersymmetric gauge theory,''
  Nucl.\ Phys.\  B {\bf 448}, 93 (1995)
  [arXiv:hep-th/9505062].
  %%CITATION = NUPHA,B448,93;%%
}

\lref\WittenQJ{
  E.~Witten,
  ``Anti-de Sitter space and holography,''
  Adv.\ Theor.\ Math.\ Phys.\  {\bf 2}, 253 (1998)
  [arXiv:hep-th/9802150].
  %%CITATION = 00203,2,253;%%
}
\lref\GubserBC{
  S.~S.~Gubser, I.~R.~Klebanov and A.~M.~Polyakov,
  ``Gauge theory correlators from non-critical string theory,''
  Phys.\ Lett.\  B {\bf 428}, 105 (1998)
  [arXiv:hep-th/9802109].
  %%CITATION = PHLTA,B428,105;%%
}
\lref\MaldacenaRE{
  J.~M.~Maldacena,
  ``The large $N$ limit of superconformal field theories and supergravity,''
  Adv.\ Theor.\ Math.\ Phys.\  {\bf 2}, 231 (1998)
  [Int.\ J.\ Theor.\ Phys.\  {\bf 38}, 1113 (1999)]
  [arXiv:hep-th/9711200].
  %%CITATION = IJTPB,38,1113;%%
}
\lref\GubserVD{
  S.~S.~Gubser,
  ``Einstein manifolds and conformal field theories,''
  Phys.\ Rev.\  D {\bf 59}, 025006 (1999)
  [arXiv:hep-th/9807164].
  %%CITATION = PHRVA,D59,025006;%%
}
\lref\BarnesBW{
  E.~Barnes, E.~Gorbatov, K.~Intriligator and J.~Wright,
  ``Current correlators and AdS/CFT geometry,''
  Nucl.\ Phys.\  B {\bf 732}, 89 (2006)
  [arXiv:hep-th/0507146].
  %%CITATION = NUPHA,B732,89;%%
}

\lref\DasguptaIJ{
  K.~Dasgupta and S.~Mukhi,
  ``F-theory at constant coupling,''
  Phys.\ Lett.\  B {\bf 385}, 125 (1996)
  [arXiv:hep-th/9606044].
  %%CITATION = PHLTA,B385,125;%%
}
\lref\JohansenAM{
  A.~Johansen,
  ``A comment on BPS states in F-theory in 8 dimensions,''
  Phys.\ Lett.\  B {\bf 395}, 36 (1997)
  [arXiv:hep-th/9608186].
  %%CITATION = PHLTA,B395,36;%%
}
\lref\AnselmiYS{
  D.~Anselmi, J.~Erlich, D.~Z.~Freedman and A.~A.~Johansen,
  ``Positivity constraints on anomalies in supersymmetric gauge theories,''
  Phys.\ Rev.\  D {\bf 57}, 7570 (1998)
  [arXiv:hep-th/9711035].
  %%CITATION = PHRVA,D57,7570;%%
}
\lref\OsbornQU{
  H.~Osborn,
  ``$\cN = 1$ superconformal symmetry in four-dimensional quantum field theory,''
  Annals Phys.\  {\bf 272}, 243 (1999)
  [arXiv:hep-th/9808041].
  %%CITATION = APNYA,272,243;%%
}
\lref\BilalPH{
  A.~Bilal and C.~S.~Chu,
  ``A note on the chiral anomaly in the AdS/CFT correspondence and $1/N^2$
  correction,''
  Nucl.\ Phys.\  B {\bf 562}, 181 (1999)
  [arXiv:hep-th/9907106].
  %%CITATION = NUPHA,B562,181;%%
}

\lref\HenningsonGX{
  M.~Henningson and K.~Skenderis,
  ``The holographic Weyl anomaly,''
  JHEP {\bf 9807}, 023 (1998)
  [arXiv:hep-th/9806087].
  %%CITATION = JHEPA,9807,023;%%
}

%\ArgyresTQ
\lref\ArgyresTQ{
  P.~C.~Argyres and J.~R.~Wittig,
  ``Infinite coupling duals of N=2 gauge theories and new rank 1 superconformal
  field theories,''
  JHEP {\bf 0801}, 074 (2008)
  [arXiv:0712.2028 [hep-th]].
  %%CITATION = JHEPA,0801,074;%%
}

\Title {\vbox{\hbox{WIS/20/07-NOV-DPP}}}
{\vbox{\centerline{A holographic computation of the}
\centerline{}
\centerline{
central charges of $d=4$, ${\cal N}=2$ SCFTs}} }
 \centerline{Ofer Aharony$^{1}$ and Yuji Tachikawa$^{2}$}

\bigskip
\centerline{$^{1}$Department of Particle Physics} \centerline{Weizmann Institute of Science} \centerline{Rehovot
76100, Israel}
\centerline{\tt Ofer.Aharony@weizmann.ac.il}
\smallskip
\medskip
\centerline{$^{2}$ School of Natural Sciences,} \centerline{Institute for Advanced Study} \centerline{Princeton, NJ, 08540, USA}
\centerline{\tt yujitach@ias.edu}
\medskip
\bigskip
\medskip\noindent We use the AdS/CFT correspondence to compute the central charges
of the $d=4$, ${\cal N}=2$ superconformal field theories arising from
$N$ D3-branes at singularities in F-theory. These include the conformal theories
with $E_n$ global symmetries. We compute the central charges $a$ and $c$
related to the conformal anomaly, and also the central charges $k$ associated to the
global symmetry in these theories. All of these are related to the coefficients
of Chern-Simons terms in the dual string theory on $AdS_5$. Our computation is
exact for all values of $N$, enabling several tests of the dualities recently
proposed by Argyres and Seiberg for the $E_6$ and $E_7$ theories with $N=1$.

\Date{November 2007}

\newsec{Introduction}

Four dimensional ${\cal N}=2$ superconformal field theories (SCFTs) fall into two classes.
In one class the gauge coupling is exactly marginal (this is believed to happen in any ${\cal N}=2$ supersymmetric
gauge theory
with a vanishing one-loop beta function), giving rise to a family of SCFTs which become
weakly coupled in some limit (see, e.g., \SeibergAJ). The second class involves isolated theories with no
exactly marginal deformations. These theories cannot be studied perturbatively. In some
cases \refs{\ArgyresJJ,\ArgyresXN,\EguchiVU} one can flow to them from an asymptotically free
gauge theory, while in other cases their existence can (so far) only be derived as a low-energy limit
of the theory living on some branes in string theory.

A simple example is the low-energy
theory living on $N$ D3-branes in F-theory \refs{\VafaXN,\MorrisonNA,\MorrisonPP}
compactified on K3, which sit at an F-theory
singularity for which the dilaton is constant \refs{\SenVD,\BanksNJ,\DasguptaIJ,\MinahanFG,\MinahanCJ}.
There are seven types of such singularities,
denoted by $H_0,H_1,H_2,D_4,E_6,E_7,E_8$ according to the low-energy gauge theory on the
singularity (which is $A_n$ for the $H_n$-type singularities); this is a global symmetry in
the SCFT on the D3-branes (the global symmetry of these theories contains in addition an
$SU(2)_L$ symmetry, and an $SU(2)_R\times U(1)_R$ R-symmetry). The singularities may be viewed
as groups of 7-branes, which we will call $G$-type 7-branes for the $G$-type singularity; they
are invariant under the $SL(2,\bZ)$ duality of type IIB string theory. The SCFT of $N$ D3-branes
has an $N$-dimensional Coulomb branch (which can be visualized as motions of the D3-branes
away from the singularity), as well as a non-trivial Higgs branch (equal to the moduli space
of $N$ instantons in the group $G$).
In the $D_4$ case the string coupling is arbitrary, leading to an
SCFT with an exactly marginal deformation, which is simply the $USp(2N)$ gauge theory coupled
to a hypermultiplet in the anti-symmetric representation and four hypermultiplets in the
fundamental representation \refs{\SenVD,\BanksNJ,\AharonyEN,\DouglasJS}.
For the other singularities the string coupling is frozen, and the
corresponding theory cannot be studied perturbatively.  Interest in these theories was recently
revived by the work of Argyres and Seiberg \ArgyresCN\ which noted that the $N=1$ $E_6$ and $E_7$
theories appeared as subsectors of
strong coupling limits of other ${\cal N}=2$ SCFTs with an exactly marginal
coupling, providing a Lagrangian formulation for these theories.

${\cal N}=2$ SCFTs may be characterized by several central charges, which are related to
global symmetry ``anomalies''. For any ${\cal N}=1$ SCFT, the conformal anomaly
involves two central charges $a$ and $c$, which are related by supersymmetry
to the 't Hooft  ``anomalies'' of the
$U(1)_R$ current \refs{\AnselmiAM,\AnselmiYS}.
The two-point function of the currents of the global symmetry $G$
defines another central charge $k_G$, which by supersymmetry is related to the
``anomalous'' 3-point function involving one $U(1)_R$ current and two global symmetry
currents. These central charges are independent of exactly marginal deformations, so in
theories that have such deformations they may easily be computed in the free field theory limit.
On the other hand, it is not clear how to compute them in theories which do not have a
weak coupling limit.

In this paper we use the AdS/CFT correspondence \refs{\MaldacenaRE,\GubserBC,\WittenQJ}
to compute the central charges of all the ${\cal N}=2$ SCFTs
coming from $N$ D3-branes in F-theory. The AdS/CFT dual of these theories was constructed in
\refs{\FayyazuddinFB,\AharonyXZ} and it involves type IIB string theory on $AdS_5\times S^5$,
with $N$ units of 5-form flux on the $S^5$, and
with $G$-type 7-branes wrapped on an $S^3$ inside the $S^5$, leading to some deficit angle in
the circle surrounding them. In these theories the leading order contribution to (some of) the central
charges is of order $N^2$ and comes from the bulk \refs{\HenningsonGX,\GubserVD}, while a correction of order $N$ comes from
the effective theory living on the singularity. The central charges may either be computed directly
from their definition,
or through their supersymmetric relation to anomalous 3-point couplings of currents, which map to
coefficients of Chern-Simons terms in the low-energy supergravity on $AdS_5$.
For a specific central charge of the
$D_4$ theory the order $N$ correction was computed in \AharonyRZ, and here we generalize this computation
to all singularity types and to all central charges (though here we do not bother to carefully normalize
this contribution, but rather we use the $D_4$ case to set our normalization). In addition, we
provide arguments for the
value of the order one contribution to the central charges, leading to an exact formula for the
central charges for all values of $N$ and $G$.\foot{
Note added in v4:
The authors learned that $k_{E_8}$ for $N=1$ 
was calculated  previously in \CheungID\ 
both by field-theoretical and by string-theoretical methods.
Our result agrees with theirs.
}

As a first application of our result, we test the dualities proposed in \ArgyresCN. In that
paper the central charge $k_G$ of the $N=1$ $E_6$ and $E_7$ SCFTs was computed assuming the
duality, and it was verified that this leads to consistent results for other computations.
We compute this central charge holographically,
and obtain the same result, thus adding another
test for each of the dualities of \ArgyresCN. In addition, our computations of $a$ and $c$
for these theories provide further tests of these dualities, which are again successful.
We hope that our results will be useful for the construction of new dualities involving
these theories, and will assist in the classification of ${\cal N}=2$ SCFTs.

We begin in section 2 with a review of central charges in SCFTs, their relation to anomalies, and how they are
computed using the AdS/CFT correspondence. In section 3 we review how the $\cN=2$ SCFTs arise
from D3-branes near singularities in F-theory. In section 4 we review the AdS/CFT dual of
these SCFTs, and use this to compute their central charges and anomalies. We end in section 5
with a summary of our results, and of how they lead to tests of the dualities of \ArgyresCN.

\newsec{Central charges, anomalies and AdS/CFT}

Let us first recall the definitions of central charges in four dimensional
SCFTs, and their relation to 't Hooft global symmetry ``anomalies''
(see, e.g., \refs{\AnselmiYS,\OsbornQU}).
In a conformal theory, correlation functions of conserved currents are
highly constrained.  Two-point functions of currents $J^a_\mu $ of
a global symmetry group $G$ behave for small $x$ as
\eqn\jtwop{J^a_\mu(x) J^b_\nu(0) =
{3k_G\over 4\pi^4}\delta^{ab}{x^2g_{\mu\nu}-2x_\mu x_\nu \over x^8} +\cdots.}
We call $k_G$ the central charge of the $G$-currents, normalizing it
(as in \ArgyresCN) such that a Weyl spinor in the fundamental
representation of $SU(2)$ contributes $1$ to it.

The central charges contained in correlation functions of the energy-momentum
tensor are encoded in its response to a weakly coupled external metric, the
conformal anomaly
\eqn\tmm{
T^\mu_\mu={c\over 16\pi^2}({\rm Weyl})^2 - {a\over 16\pi^2}({\rm Euler}),}
where
\eqn\confanom{\eqalign{
({\rm Weyl})^2&= R^2_{\mu\nu\rho\sigma}-2R^2_{\mu\nu}+{1\over 3} R^2,\cr
({\rm Euler})&= R^2_{\mu\nu\rho\sigma}-4R^2_{\mu\nu}+ R^2.}}
$a$ and $c$ can be thought of as measures of the number of degrees of freedom
of the theory.

If the theory is superconformal, these central charges are relatively easy to determine,
because they are related to 't Hooft anomalies involving the global R-symmetries.
For $\cN=1$ SCFTs, $a$ and $c$ are given by \AnselmiYS
\eqn\AandC{
a={3\over32}\left[3\tr (R_{\cN=1}^3) -\tr (R_{\cN=1})\right],\qquad
c={1\over32}\left[9\tr (R_{\cN=1}^3) -5\tr (R_{\cN=1})\right].\qquad
}  Here, $\tr (R_{\cN=1}^3)$ and $\tr (R_{\cN=1})$ denote the strength
of the $U(1)_{R,\cN=1}^3$ and $U(1)_{R,\cN=1}$-gravity-gravity anomalies, normalized
so that they are given by the traces over the labels of Weyl fermions for
weakly coupled theories.
The R-symmetry of $\cN=2$ SCFTs is $U(1)_R \times SU(2)_R$,
and the  $U(1)_R$ symmetry of its $\cN=1$ subalgebra is
\eqn\noner{R_{\cN=1}=R_{\cN=2}/3+4 I_3/3,}
where  $I_a$ ($a=1,2,3$) are the generators of $SU(2)_R$.

Supersymmetry relates  the central charge $k_G$ for a flavor
symmetry $G$ to the 't Hooft anomaly via the relation  \eqn\K{
k_G\delta^{ab}=-2\tr (R_{\cN=2} T^a T^b).
} We normalize the generators $T^a$ of $G$ so that they have eigenvalues $\pm1$
when acting on the adjoint representation.
Similarly one can consider the central charges $k_{U(1)_{R,\cN=2}}$ and $k_{SU(2)_R}$ of the R-symmetries.
They are known to be directly proportional to $c$
(they sit in the same supermultiplet with the energy momentum tensor),
and are given by\eqn\C{
k_{U(1)_{R,\cN=2}}=16c,\qquad k_{SU(2)_R}=2c.
}

Let us now suppose that the SCFT has a gravitational dual defined on $AdS_5$.
The conserved currents $J_a^\mu$ corresponding to the charges
$Q_a$ are related to gauge fields $A^a_\mu$ living on $AdS_5$.
The AdS/CFT correspondence equates the bulk action with
the partition function of the SCFT with
the boundary coupling $\int A^a_\mu J_a^\mu$.
This shows a dependence on the gauge  of $A^a$ from the
cubic 't Hooft anomalies, which is reproduced by the
Chern-Simons coupling in the bulk \WittenQJ
\eqn\cscoup{
\propto \bigl(\tr (Q_a Q_b Q_c)\bigr) \int_{AdS_5} \left[A^a \wedge F^b \wedge F^c+\cdots\right],}
where $\cdots$ denotes the covariantization needed for non-Abelian symmetries.
The Chern-Simons interaction is gauge-invariant up to a boundary term,
and so the  bulk action depends on the gauge of $A^a$ at the boundary.
In a similar manner, the $U(1)$-gravity-gravity anomaly
is represented by the mixed gauge-gravity Chern-Simons interaction
\eqn\cscouptwo{\propto \bigl(\tr (Q_a)\bigr) \int_{AdS_5} A^a \wedge \tr (R \wedge R).}
Here, $R$ is the curvature two-form constructed from the five-dimensional metric.
Therefore, the calculation of the central charges using the AdS/CFT
correspondence reduces
to the determination of the Chern-Simons couplings in the dual theory on $AdS_5$.

\newsec{F-theory singularities and $\cN=2$ SCFTs}
\tablabel\sevenbranes
One fruitful way to study supersymmetric field theories in string theory
is to realize them as low-energy effective actions on D3-branes probing other branes.
To obtain an $\cN=2$ theory, we put D3-branes close to an F-theory
singularity,
which  is characterized by the monodromy of the axiodilaton it produces.
To have a superconformal theory, the dilaton gradient close to the singularity
must vanish, because this is related to the running coupling constant of the
D3-branes\foot{More precisely, it is related to the effective coupling on
the Coulomb branch, but in ${\cal N}=2$ supersymmetric theories this is related
to the beta function.}.
Ordinary $(p,q)$ 7-branes produce logarithmic running of the dilaton,
but when one combines several mutually non-local $(p,q)$ 7-branes
and puts them at one point, the dilaton can be constant.

\topinsert
\begintable
$G$ |  $H_0$ & $H_1$ & $H_2$  & $D_4$ & $E_6$ & $E_7$ & $E_8$ \cr
$n_7$ | 2  & 3  & 4 &  6  &  8 & 9 & 10  \nr
$\Delta$ | 6/5 & 4/3 & 3/2 & 2 & 3 & 4 & 6
\endtable
\centerline{\it Table \sevenbranes : Properties of F-theory singularities.}
\endinsert
The possible singularities with a constant dilaton were classified \MorrisonNA\ using
the Kodaira classification, and are tabulated in Table \sevenbranes.
There, $G$ stands for the low-energy gauge symmetry living on the singularity \refs{\DasguptaIJ,\JohansenAM},
and $n_7$ is the number of 7-branes (of different types)
used to construct the singularity.
7-branes are heavy, codimension-2 objects,
and they produce  conical singularities  in the transverse space.
The total deficit angle is proportional to the number of 7-branes $n_7$.
For convenience we parametrize the deficit angle by the change in the periodicity
of the angular coordinate around the 7-branes,
\eqn\deficit{2\pi \to 2\pi/\Delta.}
$\Delta$ is related to the number of 7-branes by
\eqn\DeltaAndN{
\Delta={12\over 12-n_7}.
}
Note that twelve $(p,q)$ 7-branes produce a deficit angle of $2\pi$, closing
up the space, while the $D_4$ singularity may be viewed as an orientifold (reflecting
the transverse space) together
with four D7-branes, so it has a deficit angle of $\pi$.

Now let us introduce $N$ D3-branes on top of the $G$-type 7-brane. The field theory
realized on the D3-branes has global symmetries coming from
the isometry of the system,
\eqn\globalsym{U(1)\times SO(4)\simeq U(1)_R \times SU(2)_R\times SU(2)_L,}
and also the flavor symmetry $G$ coming from the gauge symmetry of the 7-branes.
The theory is of rank $N$, which means that its Coulomb branch is of complex
dimension $N$. It corresponds to the motion of the $N$ D3-branes in
the transverse space to the singularity.
If we put the theory at the origin of the Coulomb branch, i.e.~if we put
all of the D3-branes together on top of the singularity,
the theory becomes conformal. One can show that the lowest dimension operator
parametrizing the position on the Coulomb branch has
dimension $\Delta$,
which explains our usage of the symbol $\Delta$ for the deficit angle. There is
also a Higgs branch emanating from the origin of the Coulomb branch, which will
not be relevant for our discussion here.

The only singularity where the value of the dilaton can be arbitrarily tuned is
the $D_4$ singularity, which in the perturbative region may be viewed as
a system of four D7-branes
on top of an $O7^-$ orientifold plane.  The gauge theory on $N$ D3-branes close to it
can be found by quantizing the open string, and is an $\cN=2$ $USp(2N)$
gauge theory with eight half-hypermultiplets in the fundamental representation ${\bf 2N}$,
and two half-hypermultiplets in the antisymmetric tensor ${\bf N(2N-1)}$.
The half-hypermultiplets in ${\bf 2N}$ transform as the fundamental of
the $D_4=SO(8)$ flavor  symmetry, and the half-hypermultiplets in ${\bf N(2N-1)}$
as the fundamental of $SU(2)_L$.
The trace part of the antisymmetric tensor corresponds to the overall motion
of the D3-branes parallel to the 7-brane, which is completely decoupled from the
rest of the theory.

The monodromy around other types of 7-branes fixes the dilaton,
so these systems are inherently strongly coupled.
The theories on  one D3-brane near the 7-brane  of type $H_{n}$
correspond to the $\cN=2$ SCFTs  found in \refs{\ArgyresJJ, \ArgyresXN},
and can be realized field theoretically
by tuning the masses and the vacuum expectation value
of an $SU(2)$ gauge theory with $n+1$ massive fundamental flavors.
The superconformal R-symmetry is an accidental symmetry at the infrared
fixed point, so  we do not have any effective means of calculating the 't Hooft anomalies
in the language of the four dimensional field theory.
For the 7-brane of type $E_n$, the corresponding SCFTs on a single D3-brane
were first discussed in \refs{\BanksNJ,\DasguptaIJ} and  were studied shortly thereafter
in detail by \refs{\MinahanFG,\MinahanCJ}.
A purely field-theoretical construction of these theories was not known until quite recently
when the paper \ArgyresCN\  appeared, which motivated us to re-analyze these theories.

\newsec{Anomalies and central charges from the AdS description}

Let us determine the central charges of the SCFTs described
in the previous section via the AdS/CFT correspondence.
The AdS/CFT dual of these theories was constructed in
\refs{\FayyazuddinFB,\AharonyXZ},
by taking the near-horizon limit of $N$ D3-branes sitting on
the 7-brane.
This gives type IIB string theory on
$AdS_5\times X_\Delta$, with $N$ units of $F_5$ flux.
Here $X_\Delta$ is the round 5-sphere
\eqn\sphere{
\{|x|^2+|y|^2+|z|^2={\rm constant}  \} \subset \bC^3,}
with the phase of $z$ restricted to $[0,2\pi/\Delta]$
and periodically identified. The 7-brane of type $G$ wraps
the locus $z=0$, which is a round $S^3$.
Massless gauge fields on $AdS_5$ arise both from the
isometries of $X_\Delta$, and from the gauge fields on the 7-brane.
Our objective is to find their Chern-Simons interactions, or equivalently,
the conformal anomaly coefficients $a$ and $c$
and the central charges $k_G$ and $k_L$
of the flavor symmetries $G$ and $SU(2)_L$, respectively.
We discuss the $\cO(N^2)$, $\cO(N)$ and $\cO(1)$ contributions in this order,
and denote the $\cO(N^2)$ contribution to $a$ by $a^{(2)}$, etc.

\subsec{$\cO(N^2)$ contributions}

The $\cO(N^2)$ contributions to the anomalies come only from the gravity in the bulk,
since the action of the 7-brane is of order $N$ \AharonyRZ.
The conformal anomalies
$a$ and $c$ were determined by \HenningsonGX\ to be
\eqn\leadingac{
a=c={N^2 \pi^3\over 4 \vol (X_5)}}
for general Einstein manifolds $X_5$, where $\vol (X_5)$ is the volume
of $X_5$, normalized to have a unit radius of curvature.

The scaling with $N$ and with the volume of all terms in the classical bulk action
on $AdS_5$ is easily found on dimensional grounds \GubserVD.
Let us use the coordinate system on $AdS_5\times X_\Delta$ which is
independent of the radius of curvature $R_{AdS}$ of $AdS_5$,
and put the factor $R_{AdS}^2$ in the metric.
Then, the ten dimensional Lagrangian density scales
in Planck units as $R_{AdS}^8$.
We have $N$ units of 5-form flux on
$X_5$, so $N$ is related to this by $N \propto R_{AdS}^4 \vol(X_5)$.
Thus, the Lagrangian density of the ten-dimensional bulk scales as $(N / \vol(X_5))^2$.
Integrating it over $X_5$ gives the five-dimensional
Lagrangian density scaling as $N^2 / \vol(X_5)$, which is reflected in the scaling
in \leadingac.
In our case the volume of $X_\Delta$
is that of a 5-sphere divided by $\Delta$, so we get
\eqn\AandCNsq{
a^{(2)}=c^{(2)}={N^2 \Delta\over 4}
}   to this order.

As mentioned above, the same central charges correspond to the Chern-Simons interactions
among the R-symmetry gauge fields, so they
can also be derived by decomposing
the $SU(4)^3_R$ Chern-Simons interactions of
type IIB supergravity on $AdS_5\times S^5$,
because $X_\Delta$ is locally the same as $S^5$.
This decomposition includes also the
$U(1)_R SU(2)_L^2$ Chern-Simons term, whose strength
has a fixed ratio relative to the strength of the $(U(1)_{R,\cN=1})^3$ Chern-Simons
term (which is also in the decomposition).
The latter R-symmetry anomaly is fixed by the superconformal algebra \AandC.
Thus, we get the $\cO(N^2)$ contribution to $k_L$
\eqn\KLNsq{k_L^{(2)}= {{N^2\Delta}},
}
where we used the relation \K\  between the central charge and the anomaly.
Finally, there is obviously no bulk contribution to the Chern-Simons term of
the $G$ flavor symmetry, which lives only on the 7-brane. In the $D_4$ case it
is easy to compute all of these central charges in the free field theory limit,
leading to the same results (at leading order in $1/N$).

\subsec{$\cO(N)$ contributions}

In the bulk there are no $\cO(N)$ contributions to the central charges
and anomalies, since the one-loop corrections in the bulk are $\cO(1)$. Thus, the
$\cO(N)$ contributions to the anomalies come purely from the Chern-Simons
interactions on the 7-brane at the singularity,
which include terms of the form $C_4\wedge \tr (R\wedge R)$
and $C_4\wedge \tr (F\wedge F)$.  The dimensional reduction
of these terms gives rise to five-dimensional Chern-Simons interactions,
since the five dimensional gauge fields involving the isometries include
\KimEZ, in addition to the ten dimensional metric,
a contribution of the form
$C_4\sim A_R \wedge \omega$ where $A_R$ is the $U(1)_R$ gauge potential
on $AdS_5$ and $\omega$
is the volume form of the 3-cycle wrapped by the singularity.

To determine the terms on the 7-brane,
let us first recall that the 7-brane of type $D_4$ can be realized perturbatively
as 4 D7-branes put on top of the O7$^-$-plane. The Chern-Simons
coupling on the worldvolume to the four-form field $C_4$
was determined in \AharonyRZ.
The other types of 7-branes have the dilaton pinned down to the strong coupling
region, but their Chern-Simons terms are related to the anomaly inflow
and can still be reliably determined.  Each constituent $(p,q)$ 7-brane
carries the same coupling to $C_4\wedge \tr (R\wedge R)$ (recall that all $(p,q)$
7-branes are related by the $SL(2,\bZ)$ duality symmetry which leaves $C_4$ invariant),
so when we bring
several 7-branes together the strength
of that term is proportional to $n_7$. As for the coupling
$C_4\wedge \tr (F\wedge F)$, the gauge symmetries on the worldvolume
for various types of 7-branes are related by the removal of $(p,q)$ 7-branes
one by one which enables flows between the different theories,
with a natural embedding of the (simply laced) symmetries
\eqn\embed{
A_1\subset A_2 \subset D_4 \subset E_6\subset E_7\subset E_8.}
Therefore, the strength of the coupling does not depend on the type
of the 7-brane, as long as we use the same normalization of the root vectors.
We conclude that the Chern-Simons terms on the 7-brane worldvolume are of the form
\eqn\WZ{
A\, n_7\int C_4\wedge \left[\tr (R_T\wedge R_T)-\tr (R_N\wedge R_N)\right]
+B \int C_4\wedge \tr (F\wedge F),
} with  constants $A$ and $B$ independent of the type of the 7-brane.
Here $R_{T,N}$ are the curvature of the tangent bundle and the normal bundle
of the 7-brane,
and we normalize the trace of the flavor symmetry
so that $\tr(T^a T^b) = \delta^{ab}/2$ independent of the group.

These terms reduce to
various Chern-Simons interactions in five dimensions
after the integral over the $S^3$ wrapped by the 7-brane.
The $C_4\wedge\tr (F\wedge F)$ term gives rise to the
$U(1)_RG^2$ Chern-Simons term on $AdS_5$, while the
$C_4\wedge \tr (R_T\wedge R_T)$ term in \WZ\ produces
$U(1)_R^3$ and $U(1)_R$-gravity-gravity Chern-Simons terms.
Finally, the $C_4\wedge \tr (R_N\wedge R_N)$ term gives
$U(1)_RSU(2)_R^2$ and $U(1)_RSU(2)_L^2$ Chern-Simons interactions.
Together, the $C_4\wedge \tr(R\wedge R)$ terms contribute to the anomalies
appearing in \AandC, and thus to $a$ and $c$.

Again, we can easily determine the scaling with $N$ and
with the volume of $X_\Delta$
of the terms in the five dimensional action arising from the integration of
\WZ\ on the 3-sphere and involving the $U(1)_R$ field $A_R$. On
dimensional grounds,  the full low-energy 7-brane Lagrangian density is proportional to
$R_{AdS}^4 \sim N / \vol(X_5)$. On the other hand, the volume of
the 3-sphere which the 7-brane wraps is independent of the deficit angle $\Delta$.
Thus, the coefficients of the five dimensional terms we obtain from \WZ\ scale as
$N / \vol(X_5) \propto N \Delta$.

Therefore,  we find that the $\cO(N)$ correction to the various
anomalies scales as
\eqn\onscaling{k_G^{(1)}\propto N\Delta, \qquad
k^{(1)}_L,  a^{(1)}, c^{(1)} \propto  Nn_7\Delta.}
The coefficients can be fixed by a careful computation, but instead we will
fix them by comparing them to the perturbative
case $D_4$, for which we can easily find
\eqn\dfourresults{k_G^{(1)}=4N,\quad
k_L^{(1)}=-N,\quad
a^{(1)}=N/2,\quad
c^{(1)}=3N/4,}
from the spectrum of the gauge theory.
As tabulated in Table \sevenbranes, $\Delta=2$ and $n_7=6$ for $D_4$.
Thus, we conclude that for all the theories
\eqn\orderN{
k_G^{(1)}= 2N\Delta,\qquad
k_L^{(1)}=-{Nn_7\Delta\over 12},\quad
a^{(1)}={Nn_7\Delta\over 24},\quad
c^{(1)}={ Nn_7\Delta \over 16}.
}

\subsec{$\cO(1)$ contributions}

Relatively little is known about the $\cO(1)$ corrections to the anomaly coefficients.
For the prototypical duality between the $\cN=4$ $SU(N)$ super Yang-Mills theory
and the $AdS_5\times S^5$ background,
the $\cO(1)$ correction accounts for the difference between the gauge group
$U(N)$ and $SU(N)$. On the stack of $N$ D3-branes in a flat space, we naturally
have a $U(N)$ gauge symmetry, where the $U(1)$ part
describes the center-of-mass motion of the D3-branes in the transverse $\bR^6$.
This overall motion decouples from the rest of the dynamics
in the near-horizon limit, and it is not the part of the theory dual to type IIB string
theory on $AdS_5\times S^5$.

In type IIB string theory on $AdS_5\times S^5$, the Kaluza-Klein reduction
from ten dimensions gives the coefficient $N^2$ for the $SU(4)_R^3$ Chern-Simons
term, and the problem is how to account for the extra $(-1)$ in the coefficient of
the Chern-Simons interaction.  It was argued in \BilalPH\ that the one-loop integral
of the fermionic Kaluza-Klein towers reproduces this contribution. The reasoning
goes roughly as follows:  almost all of the states in the fermionic tower
come in pairs of an $SU(4)_R$ representation and its conjugate
so that they do not contribute to the Chern-Simons term,
but one chiral fermion in the $\bf 4$ of $SU(4)_R$ is missing in the spectrum.
This fermion is in the so-called ``doubleton'' representation of $SO(4,2)$, which
naively appears in the Kaluza-Klein reduction, but in fact corresponds to pure gauge
modes.
The extra $(-1)$ in the Chern-Simons coefficient comes from the careful one-loop integration of
this fermion, which can be identified with the fermion
in the $\cN=4$ vector multiplet  corresponding to the overall motion of the D3-branes.
Thus, the one-loop contribution gives precisely minus the contribution of a single
$\cN=4$ vector multiplet.

Another argument giving the same answer is the following. In the system of
D3-branes before taking any near-horizon limit, the global anomaly must be
canceled by some anomaly inflow (as discussed e.g. in \refs{\FreedTG,\HarveyBX} for the case of
M5-branes). However, all inflow terms involve the fields generated by the
D3-branes, so they are proportional to some positive power of $N$.\foot{In other words,
the inflow vanishes for $N=0$. We assume that there
are no bulk fields localized in the same place that the D3-branes are localized; this
argument can fail when the D3-branes sit at a fixed point of some orbifold.}
Thus, there cannot
be any terms of $\cO(1)$ (or higher orders in $1/N$) in the anomaly of the D3-branes. This anomaly
includes both the non-trivial SCFT of the D3-branes and the decoupled vector
multiplet; thus, the $\cO(1)$ terms in the non-trivial SCFT are precisely
minus the contribution of a single $\cN=4$ vector multiplet, as above.

We can easily generalize both types of arguments to our setup which is
$\cN=2$ supersymmetric. Here, the degrees of freedom which decouple in the near-horizon
limit correspond to the overall motion parallel to the 7-brane, which is described by
a free hypermultiplet which transforms in the ${\bf 2}$ of $SU(2)_L$,
and is neutral under the flavor symmetry.
Again, this free hypermultiplet is in a one-to-one correspondence with the
``doubleton'' fields on $AdS_5\times X_{\Delta}$. Thus, the arguments above suggest that
the $\cO(1)$ contributions in this case should be precisely minus those of
this free hypermultiplet, namely
\eqn\orderone{
k_G^{(0)}=0,\quad
k_L^{(0)}=-1,\quad
a^{(0)}=-1/24,\quad
c^{(0)}=-1/12,
} independently of the type of the 7-brane.
This precisely agrees with the known result in the $D_4$ case, and
we will see more evidence in the next section that this procedure is correct.

\newsec{Summary of results and tests of Argyres-Seiberg dualities}

Combining the results
obtained so far, we obtain our final equations
\eqn\final{\eqalign{
k_G&=2 N \Delta, \cr
k_L&=N^2 \Delta - N (\Delta-1) -1,\cr
a&={1\over 4} N^2 \Delta+{1\over 2}N(\Delta-1)   -{1\over 24},\cr
c&={1\over 4} N^2 \Delta+{3\over 4}N(\Delta-1)   -{1\over 12}
}}
for the central charges. Here, we used the relation \DeltaAndN\ which relates the deficit angle and the number of 7-branes
to rewrite $n_7 \Delta=12(\Delta-1)$.

\tablabel\centralcharges
Based on the arguments above we believe that these formulas are exact, so we can use
them even in the case of $N=1$.
The first thing we notice is that
$k_L$ automatically vanishes if $N=1$.
This provides us with a test of our procedure
in the last section, because
the motion of a single D3-brane along the 7-brane
is always described purely by the free hypermultiplet, and hence
the rank one SCFTs do not carry the extra flavor symmetry $SU(2)_L$.
The rest of the central charges, $k_G$, $a$ and $c$ for the rank one SCFTs are
tabulated in Table \centralcharges. Note that in the $H_0$ case there is no flavor
symmetry $G$, so the value of $k_G$ which we wrote (using the formulas above) is
not really meaningful.

Let us next compare our results to the recent field-theoretical analysis in \ArgyresCN.
There, the authors analyzed rank two $\cN=2$ supersymmetric gauge theories
whose beta function is perturbatively zero. The gauge coupling constant
of such theories is exactly marginal, so one can go to the strong coupling
limit while preserving the superconformal symmetry.  They used the
Seiberg-Witten curves encoding the effective couplings on the Coulomb branch
to study two cases, and
found that a specific strong coupling limit of these theories
is given by the isolated rank one SCFTs of type $E_{6,7}$,
weakly coupled to an $SU(2)$ theory
(with additional fundamental half-hypermultiplets in one case).
This is a new type of S-duality,
and it is also the first construction of isolated SCFTs with exceptional symmetry
which uses purely four-dimensional field theory.
Furthermore, the R-symmetry of the setup is manifest throughout the procedure,
so it is easy to calculate the conformal anomalies $a$ and $c$.

\topinsert
%\begintable
%$G$ |  $H_0$ & $H_1$ & $H_2$  & $D_4$ & $E_6$ & $E_7$ & $E_8$\cr
%$k_G$ | $12\over 5$  & $8\over3$  & 3 &  4  &  6 & 8 & 12  \nr
%$a$ | $43\over 120$ & $11\over 24$ & $7\over12$ & $23\over24$ & $41\over24$
%& $59\over24$ & $95\over24$ \nr
%$c$ | $11\over30$ & $1\over2$ & $2\over3$ & $7\over6$ & $13\over6$ & $19\over6$ & $31\over6$
%\endtable
\centerline{\epsfbox{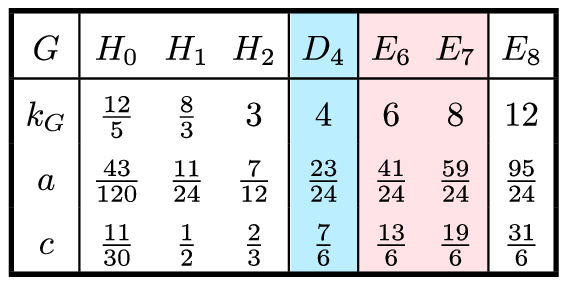}}
\centerline{\it Table \centralcharges : Central charges of rank one SCFTs. The $D_4$ case was used as an input to set}
\centerline{\it the normalizations. We found agreement with the results of \ArgyresCN\ for the $E_{6,7}$ cases.}
\endinsert

The first  case is the duality
\eqn\esix{
\hbox{$SU(3)$ with $6\times({\bf 3+\bar{\bf 3}})$ $\longleftrightarrow$
$SU(2)$ with $2\times\bf 2$ and SCFT$_{E_6}$, }
}
where $SU(2)\subset E_6$ is gauged. The consistency of this duality requires
\ArgyresCN\ the $E_6$
central charge to be $k=6$, and this matches precisely with
our calculation above using AdS/CFT. If the duality is correct, we can compute
$a$ and $c$ of the $E_6$ SCFT by
calculating those of the $SU(3)$ gauge theory
and subtracting those of the $SU(2)$ gauge theory (including the hypermultiplet
contributions) on the right hand side.\foot{
In \ArgyresCN\ the authors calculated in the same way the central charge $k_{U(1)R}$ of the
$U(1)_R$ current, which is proportional to $c$ by the relation \C.}
The results are
\eqn\esixAC{\eqalign{
a&={29\over 12}-{17\over 24}={41\over 24},\cr
c&={17\over 6}-{2\over 3}={13\over 6},
}}   which also perfectly match with our results above.

The second case is
\eqn\eseven{\hbox{$USp(4)$ with $12\times{\bf 4}$ $\longleftrightarrow$
$SU(2)$ with SCFT$_{E_7}$. }} The consistency of the duality in this case requires the
flavor current central charge to be $k=8$ \ArgyresCN.
$a$ and $c$ can be obtained just as before, with the results
\eqn\esevenAC{\eqalign{
a&={37\over 12}-{5\over 8}={59\over 24},\cr
c&={11\over 3}-{1\over 2}={19\over 6}.
}}  These also completely agree with our findings, tabulated in Table \centralcharges.

These non-trivial agreements strongly suggest that the new S-duality found
in \ArgyresCN\ and our calculation in this paper are both consistent and correct.
In Table \centralcharges, the entry for $D_4$ was used  as an input (for normalizing
the $\cO(N)$ terms),
and we compared successfully  the entries for $E_6$ and $E_7$
against the results obtained in \ArgyresCN.
The other entries are our prediction. It would be extremely interesting
to calculate $k$, $a$ and $c$ of other isolated SCFTs using some
purely field theoretical framework (such as new S-dualities), and to compare them to our findings.
Our results may be useful for pursuing that direction.
For example, the fact that $c$ of the rank one $E_8$ theory is larger than
$c$ of any rank two perturbative SCFT strongly suggests that we need to look for
SCFTs of rank more than two in order to find a dual for a theory including
the rank one $E_8$ theory. This is also suggested by the fact that since this theory
has $k_G=12$, we cannot gauge an
$SU(2)$ subgroup of $E_8$, since the resulting $SU(2)$ theory would not be
asymptotically free.

{\bf Note added in v3:}
The rank one isolated SCFT with $E_8$ flavor symmetry 
was found  in \ArgyresTQ\  as a sector of the S-dual of rank three perturbative
gauge theories, e.g. the $USp(6)$ gauge theory with ${\bf 14}+11\times {\bf 6}$.
The central charges $a,c,k_G$ for the $E_8$ SCFT calculated in their method
agreed with our prediction.

\bigskip\bigskip\bigskip
\centerline{\bf Acknowledgements}
\noindent

We would like to thank P. Argyres, B. Kol, H. Ooguri, A. Schwimmer,
N. Seiberg, S. Terashima, B. Wecht and S. Yankielowicz for useful discussions.
The work of OA is supported in part by the
Israel-U.S. Binational Science Foundation, by a center of excellence supported by the Israel Science Foundation
(grant number 1468/06), by a grant (DIP H52) of the German Israel Project Cooperation, by the European network
MRTN-CT-2004-512194, and by a grant from G.I.F., the German-Israeli Foundation for Scientific Research and
Development.
The work of YT is in part supported by the Carl and Toby Feinberg fellowship
at the Institute for Advanced Study, and by the United States
DOE Grant DE-FG02-90ER40542.

\listrefs
\bye